# The Hidden Nuclear Spectrum of the Luminous IRAS Source FSC10214+4724

Buell T. Jannuzi[1,2]

Institute for Advanced Study, Princeton, NJ 08540

email: jannuzi@guinness.ias.edu

Richard Elston

National Optical Astronomy Observatories, CTIO, Casilla 603, La Serena, Chile 1353

Gary D. Schmidt[1] and Paul S. Smith[1]

Steward Observatory, University of Arizona, Tucson, AZ 85721

H. S. Stockman

Space Telescope Science Institute[3], Baltimore, MD 21218

## ABSTRACT

Optical spectropolarimetry of the luminous IRAS source FSC10214+4724 ($z$= 2.286) reveals that the strong ($\sim$17%) linear polarization detected by Lawrence *et al.* is shared by both the narrow UV emission lines and the underlying continuum. This observation and the brightness of the source rule out synchrotron emission and dichroic extinction by dust as the polarizing mechanism, leaving scattering as the only plausible cause of the polarized emission. The narrowness of the lines requires that the scatterers be dust grains or cool ($< 1.6 \times 10^4$ K) electrons. We can recover the spectrum that is incident on the scattering medium provided we make some reasonable assumptions regarding the source geometry. The scattered UV spectrum has a power law index $\alpha$ of $-1.2 \pm 0.6$ ($F_\nu \propto \nu^\alpha$), steeper than what would be expected from a young burst of star formation, but similar to many AGN.

*Subject headings:* galaxies:individual (FSC 10214+4724) – Polarization – galaxies:peculiar – galaxies:starburst

## 1. Introduction

The large bolometric luminosity of the high-redshift IRAS source FSC10214+4724 ($z = 2.286$; $> 3 \times 10^{14}$ L$_\odot$; cf. Rowan-Robinson *et al.* 1991) has attracted the attention of observers intent on determining the mechanism(s) responsible for such a prodigious output of energy. Proposed explanations include a burst of star formation in a primeval or "proto" galaxy (Rowan-Robinson *et al.* 1993, hereafter RR93), an obscured active galactic nucleus (AGN), or some hybrid of the two (e.g. Lawrence *et al.* 1993, hereafter L93; Lawrence *et al.* 1994).

The infrared spectral energy distribution and molecular emission imply large masses of both dust ($\sim 10^9 M_\odot$; RR93) and gas ($\sim 10^{11} M_\odot$; Brown & Vanden Bout 1991, 1992ab; Solomon, Radford, & Downes 1992). The optical (rest frame UV) spectrum closely resembles a Seyfert II nucleus, and emission lines of [NII] and [OIII] are prominent (Elston *et al.* 1994, hereafter E94). Strong, narrow (FWHM $< 1200$ km/sec) H$\alpha$ emission (Soifer *et al.* 1992; E94), has been found to arise from a small but spatially resolved region ($\sim$5 kpc diameter; Matthews *et al.* 1994).

Unusual for most AGN or starburst galaxies is the large degree of linear polarization observed in the rest-UV (16.4% in a broad band; L93). Considering the brightness of the source at these wavelengths, a polarizing mechanism involving dichroic extinction by intervening dust can be effectively ruled out in favor of synchrotron emission or a scattering process. A determination of the polarizing mechanism would provide an important discriminant between competing models for FSC10214+4724.

## 2. Observations

Observations of FSC10214+4724 were obtained on 1993 April 26 UT with the CCD spectropolarimeter described by Schmidt, Stockman, & Smith (1992) mounted on the 4 m Mayall Telescope of Kitt Peak National Observatory. Five $Q - U$ sequences totaling 12000 s were obtained in moderate seeing ($\sim 1.5''$ FWHM), with the position of the object shifted along the $1.5'' \times 26''$ slit between sequences for optimum sky subtraction. The slit was placed across the object north-south. A 600 g mm$^{-1}$ grating provided spectral coverage $\lambda\lambda 4270$–7230 and resolution of 14Å FWHM. To minimize the effects of read noise, the data were binned 2×2 on-chip, resulting in pixel dimensions of $\sim 0.5''$ and $\sim 8$Å. Calibration of the polarization position angle utilized observations of standard Hiltner 960 (Schmidt, Elston, & Lupie 1992). Co-addition of the individual measurements into a final polarization spectrum incorporated the statistical quality of each $Q - U$ sequence, and a polarization uncertainty spectrum was also generated on that basis. The results are presented vs. both observed and rest wavelength in Figure 1.

Our total flux spectrum of FSC10214+4724 (Fig. 1, bottom) clearly displays the emission lines of C IV $\lambda\lambda 1548,1550$, He II $\lambda 1640$, and C III] $\lambda 1909$ together with a rather flat underlying continuum. Measurements of the wavelength of line center, equivalent width, and FWHM from a Gaussian fit to each observed profile are contained in Table 1. Also listed are line widths in km s$^{-1}$, after deconvolving the instrumental profile as measured from night sky emission lines (FWHM = 710 km s$^{-1}$). For C IV an additional correction was made for the fact that this line is an unresolved doublet. The measured C IV and C III] emission-line properties are consistent with the observations of E94, however our equivalent width for He II $\lambda 1640$ exceeds the previous report by about a factor of two. Presumably, this is the result of differences in fitting the underlying continuum, and not evidence for spectral variability.

In agreement with previous observations, the emission lines of FSC10214+4724 are narrow ($\sim 1100$ km s$^{-1}$). There is no evidence in any of the published spectroscopy (including H$\alpha$; E94) for broad components similar to those seen in a Type I Seyfert galaxy or QSO.

The spectrum of the polarized flux from FSC10214+4724 is also shown in Fig. 1.

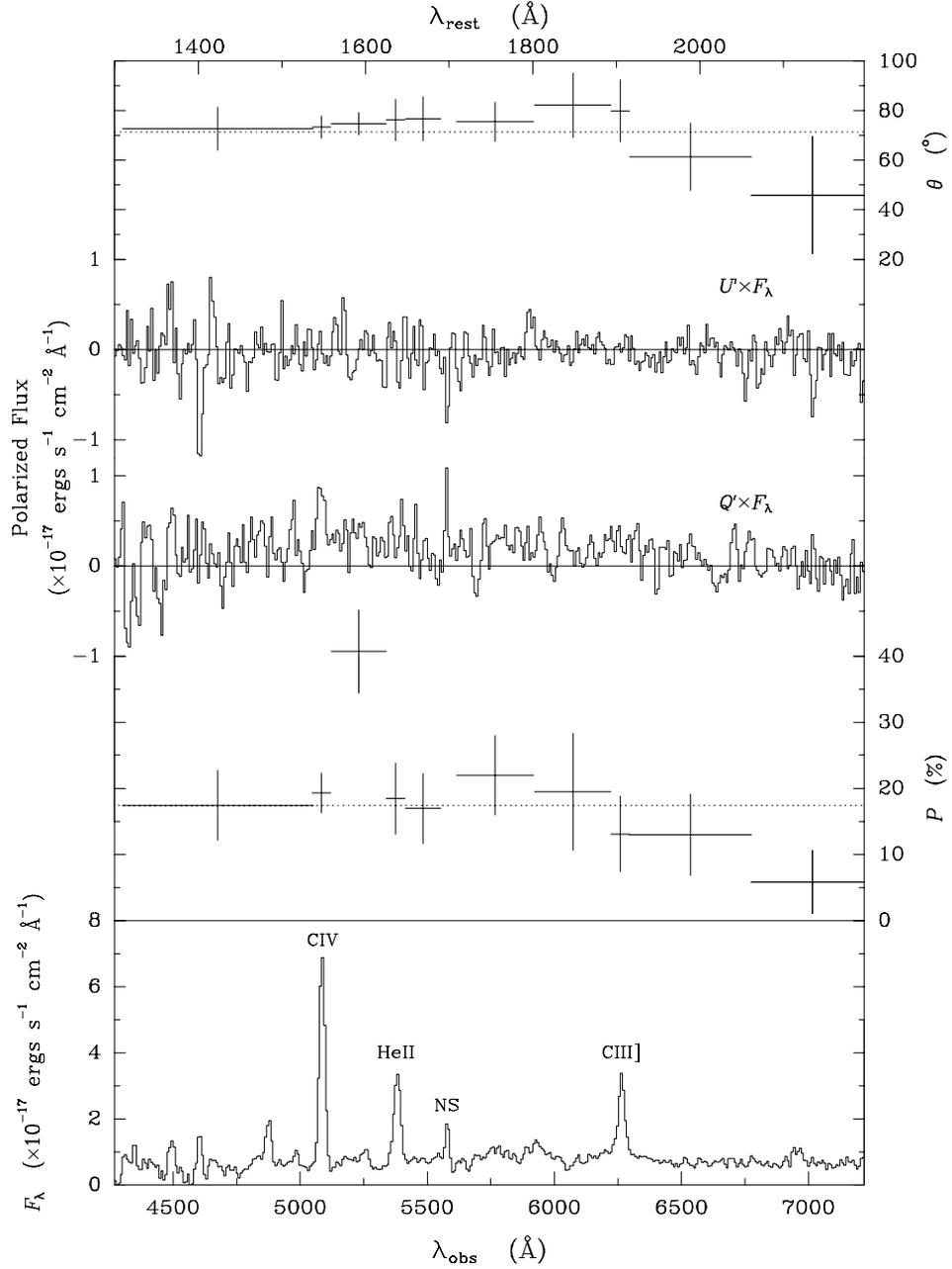

Fig. 1 – The total flux (bottom) and polarization spectra of FSC10214+4724 displayed in the observed frame. The degree ($P$) and position angle ($\theta$) of linear polarization are represented by crosses for specific wavelength bins, with horizontal bars denoting bin width and vertical bars the $1\sigma$ uncertainties in the measurement for that bin. The broad-band polarization measurement of Lawrence *et al.* (1993) is represented by the dotted lines. The spectrum of the polarized flux is displayed as the Stokes parameters $Q'$ and $U'$, which have been rotated 71° from the N-S convention in order to maximize the the polarized flux in $Q'$.

The spectrum has been rebinned from the original 8Å-wide pixels into elements chosen to isolate the emission lines and to obtain higher, roughly constant, polarimetric S/N per bin. The horizontal bars of each "cross" symbol denote the bin width and vertical bars the $\pm 1\sigma$ uncertainty as computed from the rms of values from the individual polarization measurements.

Over at least the range 1300 – 2000Å (rest frame) the emission is clearly polarized with a constant position angle (Fig. 1, top). For comparison, we have indicated with dotted lines the very broad-band ($\lambda\lambda 4000$–10000) result obtained by L93 ($P = 16.4 \pm 1.8\%$; $\theta = 75° \pm 3°$). An average over our entire spectral range (including the emission lines) yields $P = 17.1 \pm 2.9\%$; $\theta = 71°5 \pm 4°9$, in good agreement with those values.

It is also clear that the narrow emission lines do not dilute the polarization nor rotate the position angle of the underlying continuum. Values for the emission-line polarization with the continuum subtracted are given in Table 1. Consistent results are obtained for all features, with the most significant detection being $P = 19.3 \pm 3.0\%$; $\theta = 73°4 \pm 4°4$ for the C IV emission. *The narrow line emission from FSC10214+4724 is polarized to the same degree and at the same position angle as the continuum.* We also show in Figure 1 the Stokes spectra of polarized flux, $Q' \times F_\lambda$ and $U' \times F_\lambda$, in a coordinate system rotated such that the polarized flux appears entirely in the $Q'$ parameter (i.e. a rotation of 71° from the N-S convention). The C IV line is clearly visible.

## 3. The Polarizing Mechanism

A degree and position angle of polarization which are common to both the line and the continuum emission is compelling evidence for a single polarizing mechanism and rules out the synchrotron process as the source of the polarized flux. With dichroic extinction by aligned dust grains also precluded (§1, L93 and E94), scattering – by electrons or dust grains – remains as the only plausible mechanism for the production of the polarized

TABLE 1

Emission Line Data

| Line ID | $\lambda$ (Å) | $\sigma_\lambda$ (Å) | FWHM (Å) | $\sigma$ (Å) | FWHM (km/s) | $W_\lambda$ (Å) | $\sigma_W$ (Å) | $P$ (%) | $\sigma_p$ (%) | $\theta$ (°) | $\sigma_\theta$ (°) |
|---|---|---|---|---|---|---|---|---|---|---|---|
| CIV (1548,1550) | 5086.2 | 1.1 | 24.6 | 0.8 | 1149 | 390 | 43 | 19.3 | 3.0 | 73.4 | 4.4 |
| HeII (1640) | 5382.7 | 1.1 | 29.9 | 1.5 | 1497 | 220 | 32 | 18.4 | 5.4 | 76.3 | 8.4 |
| CIII] (1909) | 6264.8 | 1.3 | 28.6 | 2.8 | 1158 | 88 | 16 | 13.1 | 5.7 | 80.0 | 12.6 |

emission from FSC10214+4724.

The remainder of this paper is devoted to considering two related but separable questions: What is the scattering agent in FSC10214+4724? What is the origin of the radiation being scattered? Answers to both are crucial for understanding the immense energy output of FSC10214+4724 and for correctly interpreting the object in the general context of galaxies at high redshift.

## 3.1. Identifying the Scatterers

We immediately rule out hot ($\gtrsim 1.6 \times 10^4$ K) electrons as the scatterers due to the narrowness of the polarized UV lines. A scattering region composed of cool ($\sim 10^4$ K) electrons would be accompanied by recombination radiation, and it is useful to examine what constraints the observed strength of H$\alpha$ would place on the size and location of such a region.

The optimum configuration for polarization by scattering is an optically thin cloud displaced on the sky from a source of illuminating ("nuclear") radiation. Due to the very high net polarization of FSC10214+4724, we assume that the central source must be hidden from our direct view in the rest-UV, but the scattering cloud is more or less unobscured. The fraction of the intrinsic continuum luminosity which is scattered is then just proportional to the total solid angle subtended by particles (here electrons) at the nucleus:

$$f = \frac{L_{\nu,s}}{L_{\nu,i}} \sim \frac{n_e \, \sigma_e \, l^3}{4\pi \, s^2} \, ,$$

where $n_e$ is the electron number density, $\sigma_e$ the Thomson scattering cross-section, $l$ the characteristic size of the cloud, and $s$ the distance of the cloud from the source. The emitted H$\alpha$ luminosity from this cloud cannot exceed that observed: $\alpha_{\mathrm{H}\alpha} \, n_e^2 \, l^3 < 4\pi \, D_L^2 \, F_{\mathrm{H}\alpha}$, where $D_L$ is the luminosity distance. Combining, we obtain

$$\frac{s^4}{l^3} < \frac{\sigma_e^2 \, D_L^2 \, F_{\mathrm{H}\alpha}}{4\pi \, \alpha \, f^2} = 0.1 \, f^{-2} \, (100/H_0)^2 \; \mathrm{pc} \, ,$$

where we have used $F_{\mathrm{H}\alpha} = 2.8 \times 10^{-15}$ erg cm$^{-2}$ s$^{-1}$ (E94) and $\Omega_0 = 1$.

Assuming isotropic scattering, $f$ can be written as the ratio of the scattered continuum flux measured at the Earth, $F_{\nu,s}$, to the intrinsic flux of the nuclear source were it seen unobstructed, $F_{\nu,i}$. The minimum value for $F_{\nu,s}$ is the value of the polarized flux in the UV continuum, $P \times F_\nu(1500\text{Å}) = 2.7 \times 10^{-29}$ erg cm$^{-2}$ s$^{-1}$ Hz$^{-1}$ (Fig 1). To estimate $F_{\nu,i}$, we assume that the tremendous $10 - 300\mu$m emission of FSC10214+4724 is powered by a spectrum resembling that of the only other object with a similar total luminosity: a quasar. For the Palomar-Green sample, Sanders et al. (1989) find a mean "bolometric correction" for the rest $B$-band of $L_{bol}/\nu L_\nu(B) \sim 16.5$. With $F_{bol} \sim 1 \times 10^{-10}$ erg cm$^{-2}$ s$^{-1}$ (RR93), this implies $F_{\nu,i}(1500\text{Å}) \sim 3 \times 10^{-27}$ erg cm$^{-2}$ s$^{-1}$ Hz$^{-1}$. We obtain $f \gtrsim 0.01$, and

$$\frac{s^4}{l^3} \lesssim 1\,(100/H_0)^2 \text{ kpc}\,.$$

Our assumed scattering geometry, and the production of strong polarization, requires $s/l \gtrsim 1$. Thus, electrons as the scatterers dictate the cloud be very close to the nucleus, $s \lesssim 1$ kpc. The H$\alpha$ angular diameter of $0.''5$ (2 $(100/H_0)$ kpc) measured by Matthews et al. (1993) provides a similar limit, if electrons are indeed responsible. We point out, however, that *more than* $10^{14}$ $L_\odot$ – *including the polarized CIV emission – must be produced within this volume*. In fact, the geometry we have chosen is the *best* case for electrons as the scatterers. Any modification to a more realistic structure (e.g. a scattering torus, a scattered or nuclear component to H$\alpha$, extinction between the nuclear source and cloud) reduces the allowed ratio of emitted H$\alpha$ to scattered UV and requires the cloud to be even *closer* to the nucleus. We also note that cool, dilute gas is probably not stable to Compton heating by the X-ray flux expected from an AGN nucleus, but may exist around a starburst region.

The alternative to electrons is dust, which must be abundant in FSC10214+4724 in order to reprocess the nuclear spectrum into the powerful IR emitter which is observed. With a scattering cross-section per unit mass $\sim 10^4 \times$ that of the electron, dust can be situated at larger distances from the source and does not require accompanying line emission.

## 3.2. Recovering the Incident Spectrum

If our view of the scattering cloud is unobscured, the observed polarized flux is a record of the spectrum of light incident on the scattering particles. Since the position angle is independent of wavelength, the polarized flux spectrum is well represented by the rotated Stokes vector $Q' \times F$ shown in Fig. 1. We have chosen to characterize the shape of the scattered spectrum by a non-linear least-squares fit of the polarized continuum points to a power law ($F_\nu \propto \nu^\alpha$). Since our data is increasingly uncertain in the absolute flux calibration at short wavelengths and there are difficulties subtracting night-sky emission bands longward of 6900Å, fits were performed over various spectral intervals as well as the entire range of the data. The values of the power-law index, $\alpha$, resulting from the various fits ranged from $-1.0$ to $-1.5$, with a best estimate judged to be $\alpha = -1.2 \pm 0.6$ for the rest-frame interval 1300 to 2100Å.

With dust grains being the most likely scattering particles (§3.1), the observed spectrum is likely flatter (bluer) than that incident on the cloud. The importance of this effect varies with grain size and composition. For the wide variety of dust properties considered by Draine and Lee (1984) and Pei (1992) (and including both Galactic and SMC dust), albedos range between extremes of being roughly neutral to flattening by nearly $+2$ in the spectral index. It is therefore reasonable to bound the incident spectrum of FSC10214+4724 by the values $\alpha = -1$ to $-3$. This range is similar to that seen for BL Lacertae objects, steeper than the spectra of most quasars, but certainly steeper than the emission from a starburst, which appears nearly flat from 1000 to 2100Å (i.e. $\alpha \approx 0$; e.g. Bruzual and Charlot 1993).

A nuclear starburst model might be constructed if the light were reddened by dust to the requisite slope of $\alpha = -1$ to $-3$ before reaching the scattering cloud. However, each unitary change in the spectral slope ($\alpha$ to $\alpha - 1$) over the wavelength region 1200 to 2200 Å requires an additional 2.5 magnitudes of extinction at these wavelengths. The required reddening to change a starburst ($\alpha = 0$) to the observed value would imply an intrinsic luminosity at 1500Å which exceeds that of a starburst with the bolometric luminosity of FSC10214+4724.

## 4. AGN, Starburst, or Hybrid?

The extreme luminosity from a compact region, the Seyfert II nuclear spectrum, possible flux variability (E94), and other properties discussed above support the contention that FSC10214+4724 contains a hidden AGN. This would not be a unique occurrence. Other luminous IRAS sources, similar in many of their properties to FSC10214+4724 (including a highly polarized continuum) appear to contain obscured AGN (e.g. IRAS 09104+4109 Hines & Wills 1993). The existence of an AGN in FSC10214+4724 complicates the interpretation of this object as a "proto" galaxy. Of course, galaxy formation and AGN may be closely related phenomena. However, the data presented in this *Letter* supports the argument that the observed UV radiation is dominated by reprocessed AGN emission and not by the light from a young stellar population.

Infrared imaging (E94, Matthews *et al.* 1994) reveals galaxies in the field which, if close companions, might be triggering the feeding of an AGN and/or starburst through tidal interactions. Alternatively, given the brightness and red colors of the companion objects, they are likely to be a foreground group (E94). In fact the northern companion might well prove to be lensing an obscured high redshift Seyfert II which is seen as the arc-like FSC 10214+4724. In this case we would have to jettison all arguments based on the total luminosity and extended emission and replace them with a lensing model for the source. Fortunately, spectroscopy of the companion should settle this question.

We thank the staff of KPNO, especially Skip Andre, for providing the technical assistance necessary to ensure smooth operation of the spectropolarimeter on the Mayall 4 m telescope. We thank B. Draine, A. Laor, and D. Weinberg for comments which improved this work. This research was partially supported by NASA grant NAG 5–1630 and NSF grant AST 91–14087. B.T.J. acknowledges support from NASA through grant number HF–1045.01–93A from the Space Telescope Science Institute, which is operated by the Association of Universities for Research in Astronomy, Incorporated, under NASA contract NAS5–26555.